\documentclass[12pt,onecolumn]{prop}

\usepackage{amsmath}
\usepackage{amssymb}

\usepackage{cite}

\def\ergscm{erg~s$^{-1}$~cm$^{-2}$}

\def\***#1{\textsf{\fontseries{b}\selectfont ***#1***}}

\long\def\ITEM{\par\medskip$\bullet$~}

\def\LCDM{$\Lambda$CDM}

\usepackage{graphicx}
\usepackage{fullpage}
\begin{document}

\begin{center}
{\bf \Large X-ray Cluster Cosmology}

{\large A white paper submitted to the Decadal Survey Committee}
\end{center}
\vspace{0.2cm}
{\bf Authors:} A. Vikhlinin$^1$, S. Murray$^1$, R. Gilli$^2$, P. Tozzi$^3$, M. Paolillo$^4$, N. Brandt$^8$, G. Tagliaferri$^9$, M. Bautz$^{12}$,
S. Allen$^{13}$, M. Donahue$^{14}$, A. Evrad$^{15}$, K. Flanagan$^{16}$,
P. Rosati$^6$, S. Borgani$^5$, R. Giacconi$^{10}$, M. Weisskopf$^7$, A. Ptak$^{10}$,
D. Alexander$^{11}$, G. Pareschi$^9$, W. Forman$^1$, C. Jones$^1$\\
\noindent
{\footnotesize
1. Harvard-Smithsonian Center for Astrophysics, Cambridge MA\\
2. INAF-Osservatorio Astronomico di Bologna, Bologna, Italy\\
3. INAF-Osservatorio Astronomico di Trieste, Trieste, Italy\\
4. Universit\`a di Napoli, Napoli, Italy\\
5. University of Trieste, Trieste, Italy\\
6. European Southern Observatory (ESO), Garching bei Muenchen, Germany\\
7. NASA Marshall Space Flight Center, Huntsville AL\\
8. Penn State University, University Park PA\\
9. INAF-Osservatotio Astronomico di Brera, Milano, Italy\\
10. The Johns Hopkins University, Baltimore MD\\ 
11. University of Durham, Durham, United Kingdom\\
12. Massachuesetts Institute of Technology, Cambridge MA\\
13. Standford University, Stanford CA\\
14  Michigan State University, E. Lansing MI\\
15. University of Michigan, Ann Arbor MI\\
16. Space Telescope Science Institute, Baltimore MD\\
}
\begin{center}
Science Frontier Panels\\
Primary Panel:Cosmology and Fundamental Physics (GFP)\\
Secondary panel:  Galaxies across Cosmic Time (GCT)\\
\vspace{0.2cm}
Project emphasized: The Wide-Field X-Ray Telescope (WFXT);
http://wfxt.pha.jhu.edu/
\end{center}
\vspace{0.25in}

\begin{figure}[h]
\centerline{\includegraphics[height=0.42\linewidth]{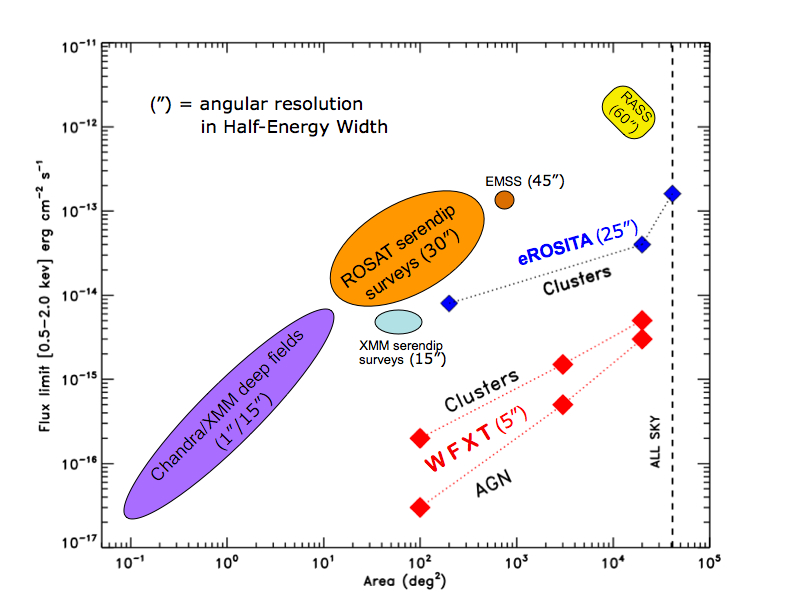}}
\end{figure}

\pagebreak

\section{Introduction}

\begin{figure}[b]
\centerline{\includegraphics[width=0.5\columnwidth]{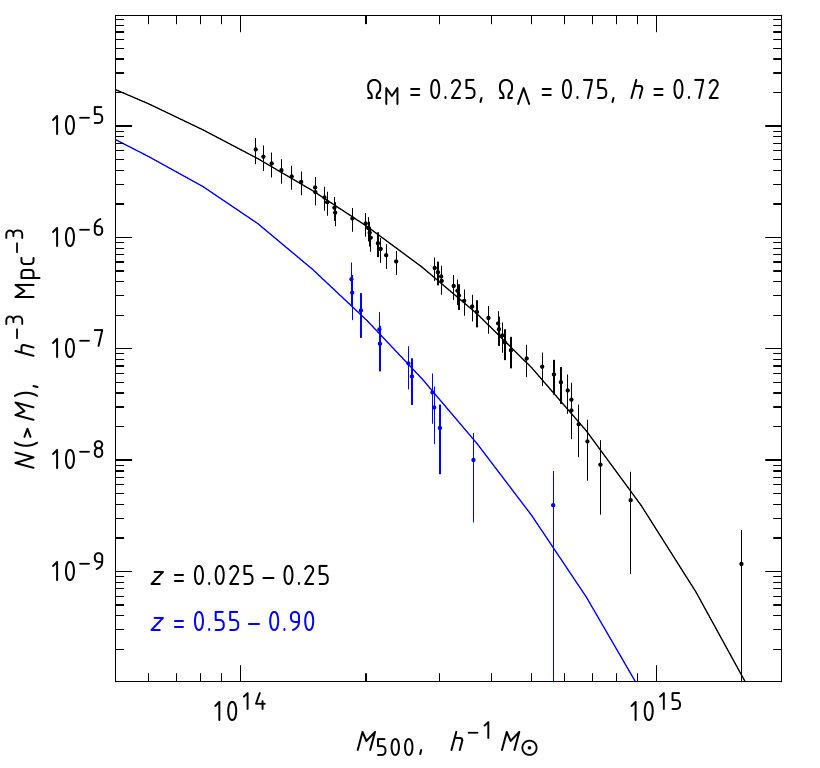}\includegraphics[width=0.5\columnwidth]{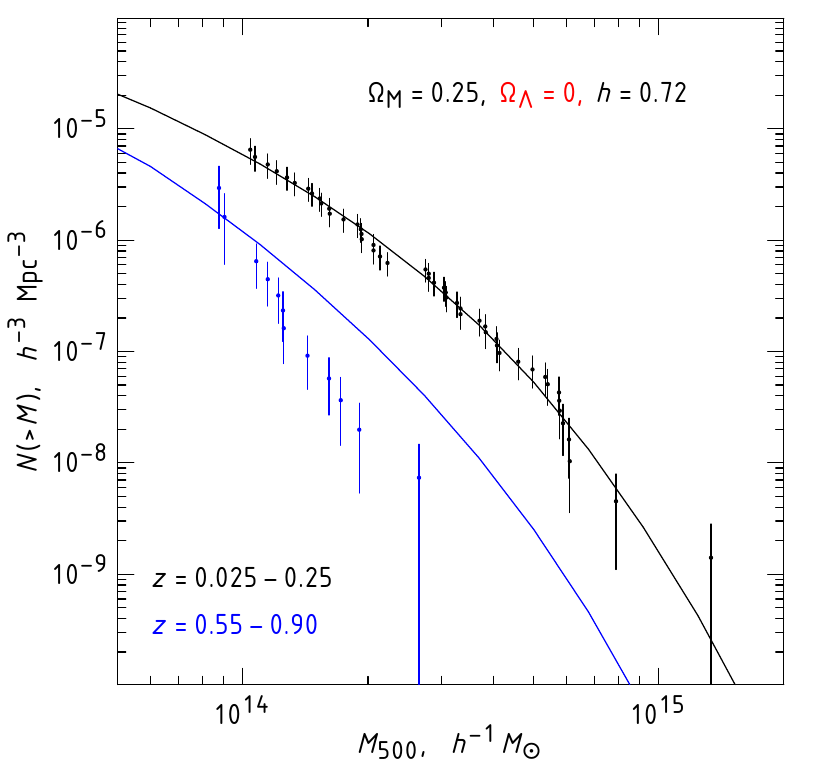}}
\caption{Illustration of sensitivity of the cluster mass function to the
  cosmological model. In the left panel, we show the measured mass
  function and predicted models (with only the overall normalization at
  $z=0$ adjusted) computed for a cosmology which is close Concordance
  $\Lambda$CDM. In the right panel, both the data and the models are
  computed for a cosmology with $\Omega_\Lambda=0$.  When the overall model
  normalization is adjusted to the low-$z$ mass function, the predicted
  number density of $z>0.55$ clusters is in strong disagreement with the
  data, and therefore this combination of $\Omega_M$ and $\Omega_\Lambda$ can be
  rejected. Reproduced from \cite{2008arXiv0812.2720V}.}
\label{fig:mfun:lambda-nolambda}
\end{figure}

Cosmological studies with galaxy clusters, and in the X-ray band in
particular, has played an important role in establishing the current
cosmological paradigm. Starting with 1990's, they have consistently
indicated low values of $\Omega_M$ (both from the baryonic fraction
arguments \cite{1993Natur.366..429W} and measurements of the evolution
in the cluster number density
\cite{1998MNRAS.298.1145E,2001ApJ...561...13B}) and low values of
$\sigma_8$
\cite{1991ApJ...372..410H,2002ApJ...567..716R,2003A&A...398..867S} --- a
result since confirmed by cosmic microwave background (CMB) studies,
cosmic shear, and other experiments
\cite{2007ApJS..170..377S,2008arXiv0803.0547K,2008arXiv0811.4280D,2007MNRAS.381..702B,2008A&A...479....9F}.
Recently, X-ray study of the evolution of the cluster mass function at
$z=0-0.8$ have convincingly demonstrated that the growth of cosmic
structure has slowed down at $z<1$ due to the effects of dark energy,
and these measurements have been used to improve the determination of
the equation state parameter \cite{2008arXiv0812.2720V}. It is
astonishing that these results are still based on samples of $\sim 100$
clusters derived from old ROSAT surveys
(Fig.\ref{fig:mfun:lambda-nolambda}). A new, sensitive, X-ray survey
should be able to make a quantum leap in the cluster cosmology science.

The main emphasis of a next-generation X-ray survey is to push the
discovery space beyond the currently achievable limits in many areas of
X-ray astronomy (AGNs, galaxies, X-ray stars, clusters and groups of
galaxies, discussed in other white papers \cite{2009WP_Murray, 2009WP_Giacconi, 2009WP_Ptak}). We would like to make
the same emphasis for cosmology. The survey will provide exceptionally
detailed information about the population of galaxy clusters and groups
over a very wide range of masses, redshifts, and spatial scales \cite{2009WP_Giacconi}. It would be possible to use these data for the
currently discussed cosmological applications such as constraints on the
Dark Energy equation of state.  However, we would like to push beyond these problems and make large X-ray survey data maximally useful for
potential discoveries such as looking for departures from the
concordance $\Lambda$CDM cosmology.

In the X-ray band, there is a huge gap in sensitivity between existing
wide-field and deep surveys. The most sensitive wide survey,
\emph{ROSAT} All-Sky Survey, is 4.5 orders of magnitude shallower than
\emph{Chandra} Deep Fields. The equivalent sensitivity gap in the
optical, between SDSS and Hubble Deep Fields, is only a factor of
250. A wide \emph{and} sensitive X-ray survey is long overdue,
especially so since all the required technology has long been in
place. In this White Paper, we consider cosmological applications of a
next-generation X-ray survey. A particular implementation we have in
mind is the Wide Field X-ray Telescope (WFXT) mission
\cite{2008SPIE.7011E..46M}, but the same considerations apply to any
experiment meeting the large effective area and high angular
resolution across a wide field of view requirements essential for
making an ``order of magnitude'' step in X-ray survey science.

The plan of this White Paper is as follows. First, we outline the use
of a large area, sensitive X-ray survey for traditional cosmological
tests, and then discuss what aspects of the cluster data can be
potentially useful for new discoveries in cosmology. We then briefly
discuss the WFXT capabilities for surveys.

\section{Traditional cluster cosmology with WFXT}

\begin{figure}[t]
  \centerline{\includegraphics[width=0.55\linewidth]{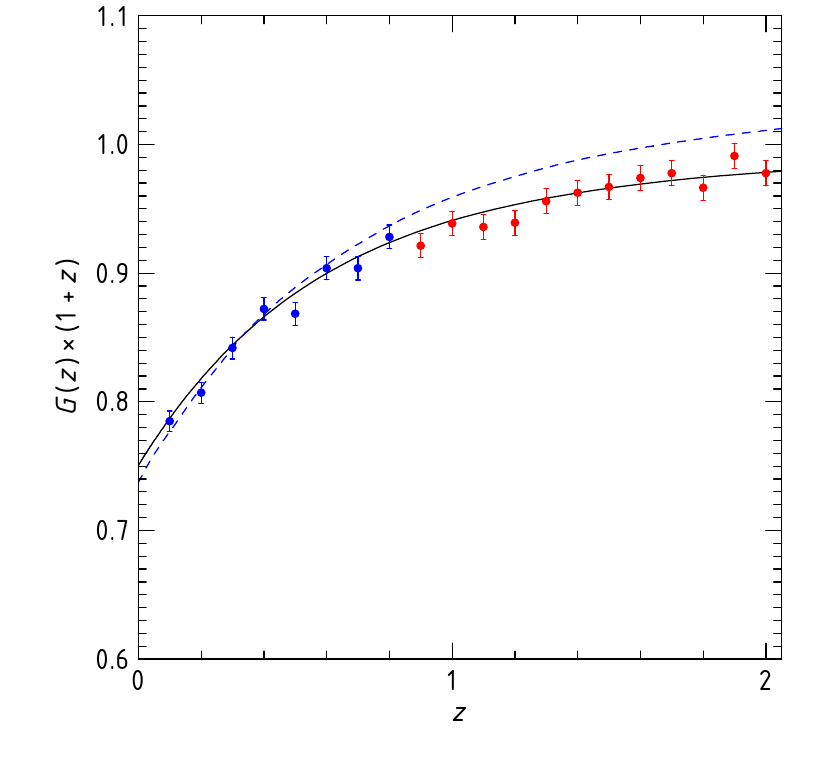}}
  \caption{The normalized growth factor of density perturbations,
    $G(z)$, constructed from X-ray and weak lensing observations of 2000
    clusters detected in sensitive X-ray surveys.  The extension of
    $G(z)$ measurements beyond $z=2.0$ will be possible only with
    pointed observations with a powerful X-ray telescope, such as IXO.
    However, at $z\lesssim 1$, the WFXT will provide sufficient
    information to carry out this measurement right from the survey data
    (see text). The growth of structure data are crucial for testing
    non-GR theories of cosmic acceleration. For example, the dashed line
    shows the $G(z)$ function predicted for a DGP model with the same
    expansion history as the quintessence model depicted by the solid
    curve.}
\label{fig:G(z)}
\end{figure}

A wide WFXT survey will provide a rich dataset for classical cosmological
tests using clusters of galaxies. The number of detected clusters will be so
large, that the linear perturbations factor can be derived with 1\%
uncertainties in each $\Delta z=0.1$ bin out to $z\simeq2$ using followup
observations with a powerful X-ray telescope (such as IXO) and weak lensing
measurements in the optical (the procedure is outlined in the White
Paper by Vikhlinin, Allen et al.). The suggested approach starts with
clusters selected in a clean survey out to $z=2$; WFXT-type survey would
be the best dataset for this purpose. A subset of the most massive
clusters ($\sim 100$ per $\Delta z=0.1$ bin) will be selected for more
detailed observations in the X-ray and optical. X-ray data from a
powerful X-ray telescope will be used to measure high-quality total mass
proxies such as $Y_X$ \cite{2006ApJ...650..128K}; the average weak
lensing signal from selected clusters is used to precisely normalize the
$Y_X-M_{\text{tot}}$ relation with minimal reliance on theory of cluster
formation. A combination of low-scatter X-ray mass proxies and bias-free
calibration of $M_{\text{tot}}$ by weak lensing leads to $\sim 1\%$
measurements of the perturbation growth factor in each redshift bin
(Fig.\ref{fig:G(z)}). Such a measurement will be extremely useful for
improving constraints on the dark energy equation of state from
geometric methods, and for testing non-GR models of cosmic
acceleration. 

The role of WFXT can go well beyond simply providing the initial cluster
sample for such an experiment. Uniquely among any other proposed survey
missions, WFXT will detect enough photons from clusters with fluxes
$f_x>3\times10^{-14}$~\ergscm{} so that high-quality proxies for the
total cluster mass will be measured. The redshift boundary of this
brighter subsample will be around $z=1$. For $z\lesssim0.8$, the
sensitivity of forthcoming optical surveys such as PanSTARRS will be
sufficient to measure the average weak lensing sheer and thus to
precisely calibrate the proxy vs.\ mass relations. Therefore, the growth
function as a function of $z$ can be measured out to $z=0.8$ without the
need for any additional observations. The growth function which can be
derived from the WFXT-only data is illustrated in Fig.\ref{fig:G(z)}
(blue). In combination with the Planck CMB priors, these data will
constrain the growth index, $\gamma$ \cite{2007PhRvD..75b3519H}, to
$\pm0.042$. This level of accuracy is best appreciated by comparison
with projections for the cosmic sheer measurements from a dedicated Dark
Energy mission (SNAP), $\Delta\gamma=\pm0.044$
\cite{2007PhRvD..75b3519H}.

For several hundred clusters at $z\lesssim 0.2$ the WFXT statistics will
be sufficient to measure radial temperature profiles of the intracluster
gas.  These clusters will be a premier, uniform dataset for
cross-correlations with the SZ data from Planck or ground-based SZ
surveys. The main application of the joint X-ray and SZ analysis is to
implement the classical method of determining the absolute distance
scale \cite{2006ApJ...647...25B}.

The WFXT sample of several hundred thousands groups and clusters will
provide an independent measurement of the baryonic acoustic oscillations
peak in the cluster-cluster power spectrum. This measurement will nicely
complement the larger-scale BAO measurements with JDEM and dedicated
ground-based galaxy surveys, because observing BAO with cluster-cluster
correlation has advantages over the galaxy-galaxy correlation
measurements due to simpler modeling of the scale-dependent bias and
possibly, of redshift-space distortions. Also, even a simple
cross-correlation of the WFXT cluster and group catalog with the sample
of luminous red galaxies\footnote{LRGs are usually considered as a prime
  tool for BAO measurements in the optical surveys; LRGs and WFXT galaxy
  clusters and groups will have a similar space density.}  will go a
long way in constraining the LRG bias factor and thus potentially
improving the accuracy of the BAO results.

Detailed sampling of the WFXT survey lightcone with galaxy clusters and
groups can also be used in cross-correlation with the CMB maps to
measure the Integrated Sachs-Wolf effect. Just as with BAO, using
clusters as tracers has advantages because of easier modeling of their
bias factor compared with normal galaxies or AGNs.

\section{Looking beyond $\Lambda$CDM}

WFXT will be able to carry out traditional cosmological tests with
galaxy clusters as outlined above. However, we believe that a more
significant contribution will be made by expanding discovery space, in
particular in looking for deviations from the ``concordant''
$\Lambda$CDM cosmological model. WFXT will contribute to this quest by
taking the cluster data of the quality and scale unimaginable before.

\ITEM Detailed measurements of the mass-dependent power spectrum of
groups and clusters, as well as the shape and evolution of the high-mass
end of the cluster mass function can be used to search for
non-gaussianities in the primordial density fluctuation field
\cite{2008PhRvD..77l3514D}. Detecting non-gaussianities would be an
extremely important discovery because it opens a window into physics of
inflation.

\ITEM WFXT will make a detailed determination of the scale, mass, and
redshift dependence of the bias factor in the cluster-cluster
correlation function. These data will be a prime dataset for testing
deviations from General Relativity on large scales. 

\ITEM A WFXT-type survey will be the first experiment to accurately
measure the cluster-cluster correlation function on mildly non-linear
scales ($\sim 1$\,Mpc)\footnote{This measurement is hard to do in any
  other waveband and even in an X-ray survey with poor angular
  resolution because of problems with separation of close neighbors.}.
This measurement is also potentially useful for testing departures from
GR (e.g., the ``chameleon'' effect in the $f(R)$ models \cite{2008arXiv0812.0545S}). Statistics of
close neighbors should be also directly linked to the growth rate of
clusters.  Therefore, comparison of the cluster-cluster correlation at
small separations and evolution in the cluster mass function will create
and extra level of ``redundant'' information on the growth of non-linear
objects, which can be exploited in testing the $\Lambda$CDM paradigm.

\ITEM Measuring the cluster-cluster power spectrum over a wide
range of spatial scales today, combined with a precise determination of
the mass-dependent bias factor, will yield a precise reconstruction of
the linear perturbations power spectrum at low redshifts. This can be
used to search for signatures of non-zero neutrino mass
\cite{2005PhST..121..153T}.

\ITEM Combining detailed information from number density and
clustering of clusters at low redshifts will lead to a very robust local
measurement of the amplitude of linear density perturbations
($\sigma_8$). A reliable \emph{absolute} measurement of the local value
of $\sigma_8$ in combination with the CMB measurements from Planck
constrains the total growth of perturbations between $z=1000$ and $z=0$.
Potential applications include searching for effects of early dark
energy, or putting tighter constraints on the ratio of tensor and scalar
perturbations in the CMB maps.

\ITEM Departures from \LCDM{} can be searched for by looking at
statistics and properties of the rarest objects, such as ``Bullet
Cluster'' \cite{2007PhRvL..98q1302F}. WFXT will dramatically extend the
search volume for such objects, especially so because it will provide
not only detections but also reasonably detailed X-ray images
($1000-1500$ photons) for virtually every massive cluster in the
surveyed lightcone out to $z\simeq 1$.

\section{\emph{WFXT} capabilities for surveys}

\begin{figure}
\centerline{\includegraphics[height=0.45\linewidth, angle=90]{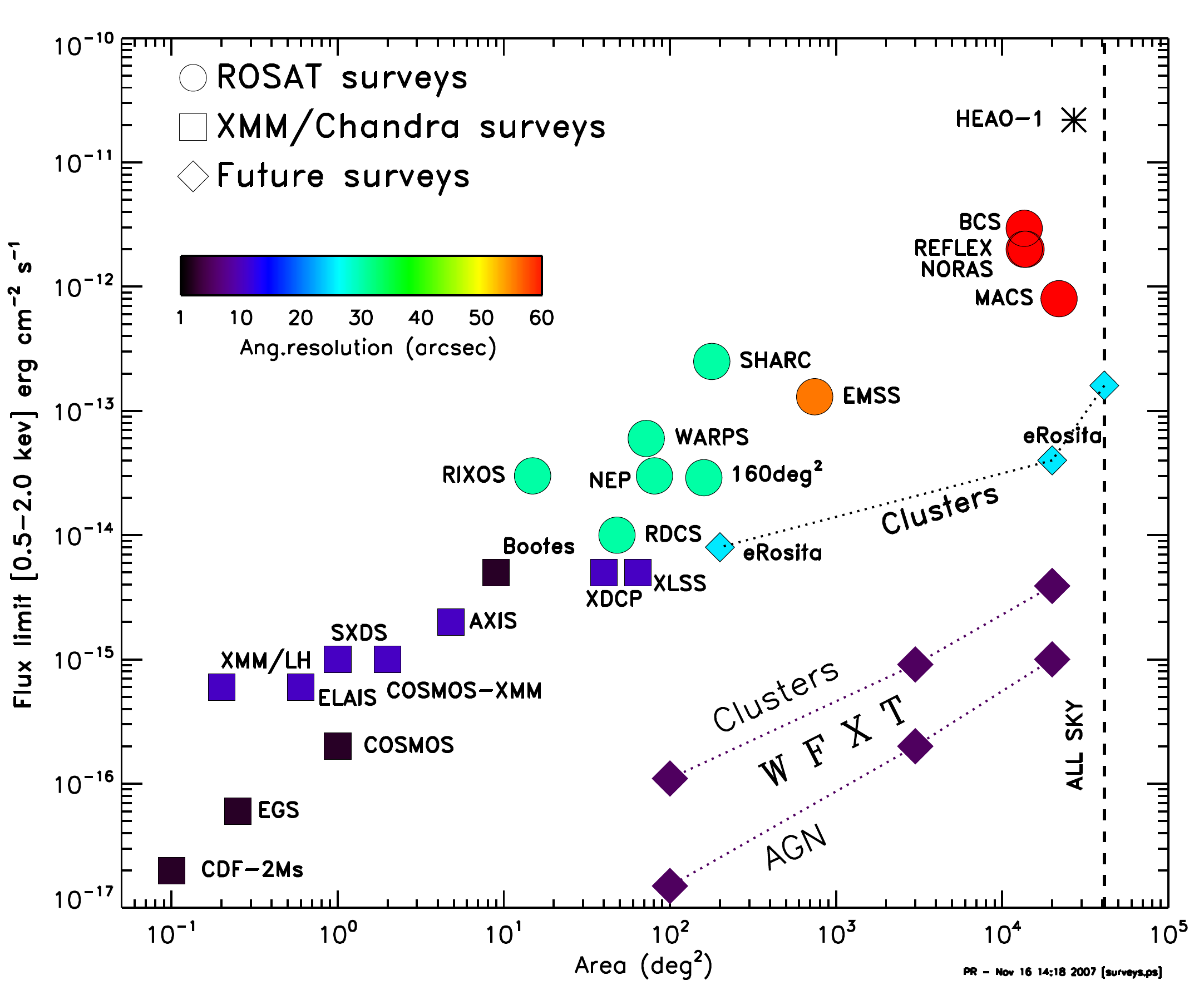}}
\caption{\footnotesize {Comparison of effective areas and sky coverage for various
  X-ray surveys, past and planned. WFXT provides an unsurpassed
  combination of sensitivity and areal coverage.}}
\label{fig:surveys}
\end{figure}

WFXT is designed to carry out an extremely advanced X-ray survey,
which will feature sensitivity limits comparable to \emph{Chandra} and
\emph{XMM-Newton} deep surveys and a sky coverage of 10\%--25\% of the
entire sky. The combination of its effective area, field of view, and
angular resolution provides unsurpassed capabilities for detection of
galaxy clusters and groups over a very broad range in mass and
redshift.  WFXT design features a total collecting area of X-ray
mirrors of 8,000~cm$^2$, in combination with an angular resolution of
$5''$ (half energy width) averaged over the entire wide field of view
of 1 deg$^2$. The high angular resolution across the field of view is
achieved using the well known polynomial pertubation to the standard
Wolter-I X-ray telescope optical design
\cite{1992ApJ...392..760B}. This design distinguishes WFXT from other
X-ray survey missions and is critical for classifying faint sources as
either point-like or extended. That is, WFXT is the analog to an
optical Schmidt telescope, optimized for large area surveys, with an
angular resolution over its entire field of view that is sufficient to
resolve extended sources and avoid confusion down to the flux limit of
the Chandra Deep Fields.  The huge grasp of the WFXT will make it
possible to carry out, in just a few years, a very sensitive X-ray
survey over a very wide area (Fig.\ref{fig:surveys}).

The WFXT survey over several thousand square degrees will reach an
extended source flux limit of $10^{-15}$~\ergscm{}. At such a low
limiting flux, the surface density of extended sources (ranging in
mass from rich clusters of galaxies at high redshifts to sub-group
objects at low~$z$) is $\sim 100$~per square degree (extrapolating the
\emph{ROSAT} $\log N-\log S$ relation and consistent with the extended
source detections by XMM in the COSMOS field).  Therefore, a WFXT
survey will be guaranteed to provide an enormous sample of clusters
and groups, several hundred thousand objects in a surveyed lightcone
spanning several thousand square degrees in the sky.

\begin{figure}[t]
\centerline{\includegraphics[height=3.0in]{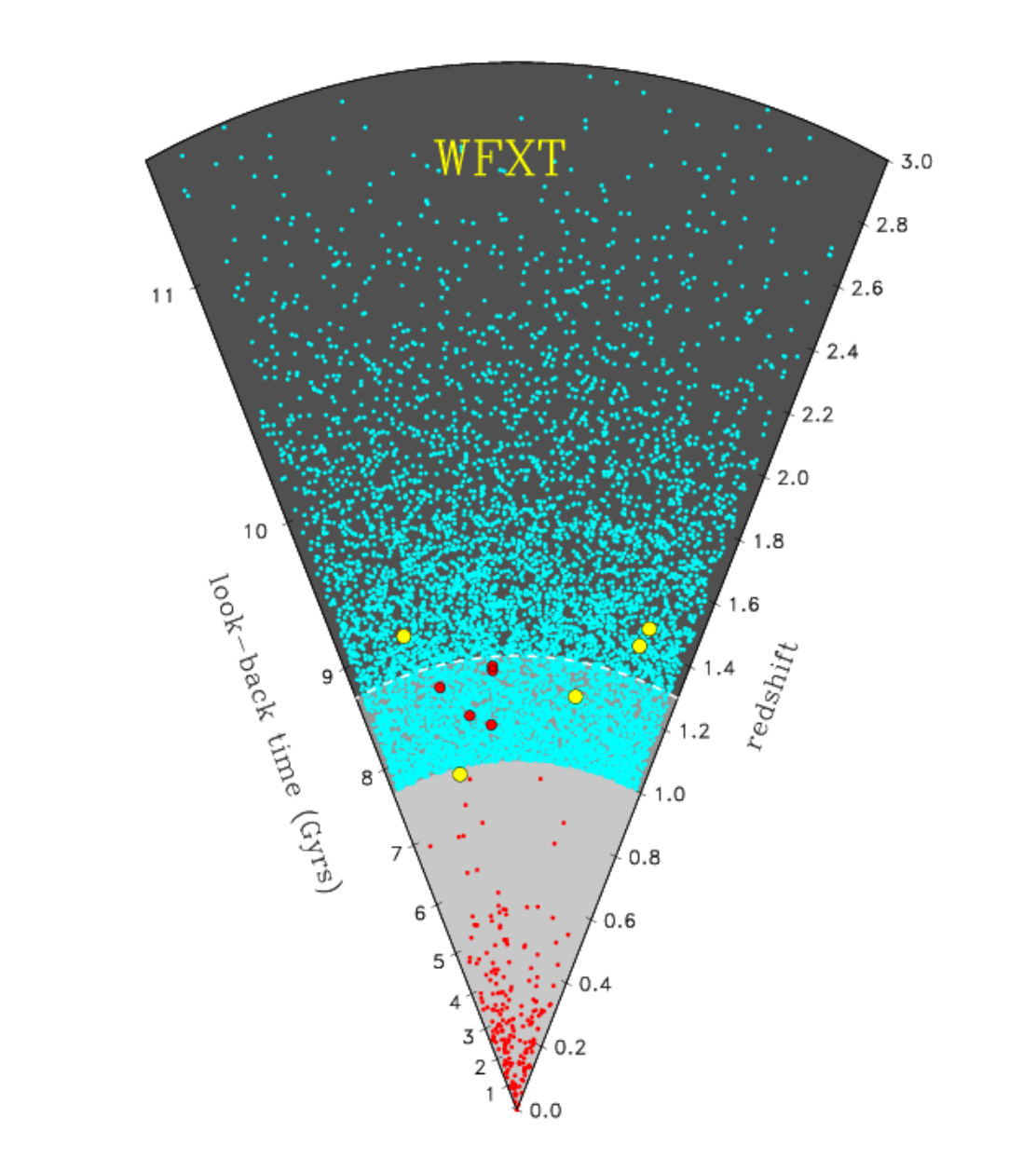}}
\caption{Look-back time cone of the expected cluster sample from the deep
  WFXT survey. WFXT will open up a largely unexplored period in cosmic
  history, back to the epoch when the first protoclusters form and the first
  virialized structures start glowing in X-rays. For clarity, we show only
  WFXT clusters at $z>1$ (blue dots). The red dots indicate clusters found
  in ROSAT deep surveys out to z=1.3; yellow dots indicate distant clusters
  found with XMM and Chandra and spectroscopically confirmed to date.}
\label{fig:lightcone}
\end{figure}

In terms of object detection, such a survey will push the currently
achievable limits in both redshift and mass. The upper boundary in
redshift for detecting massive galaxy clusters will be beyond $z=2$, the
epoch where the cluster-sized objects first came into the cosmological
scene (Fig.\ref{fig:lightcone}). WFXT will also push the mass limit at
low redshifts down to galaxy-sized objects. For example, the
``sub-groups'' with X-ray luminosities $L_X=10^{41}$~erg~s$^{-1}$ will
be detectable to $z=0.15$, increasing the search volume for such objects
by a factor of 10,000 compared with the \emph{ROSAT} All-Sky Survey and
by a factor of 30 compared even with an eRosita all-sky survey. As a
result, galaxy clusters, groups, and sub-groups detected by WFXT survey
will sample the mass function and the correlation function of the
virialized objects over an exceptionally broad range of masses,
redshifts, and spatial scales.

The WFXT X-ray telescope is so sensitive that it will provide a great
deal of detailed information for a large number of sources right from
the survey. For example, the clusters with
$f_x=3\times10^{-14}$~\ergscm{} (around the eRosita detection threshold)
--- 18,000 such objects in a 5,000~deg$^2$ survey, with an effective
redshift boundary beyond $z=1$, --- will yield 1500~photons each,
providing not only substantial information on the X-ray morphology but
also to measure the temperature, gas mass and the $Y_X$ parameter
\cite{2006ApJ...650..128K} with a 10\% accuracy. This is the same
quality of data that is presently used in the \emph{Chandra}
cosmological studies using the cluster mass function
\cite{2008arXiv0812.2720V}, only the WFXT experiment will be on a much
grander scale, using a factor of $500$ larger sample.

Many of the objects yielding 1,500 photons or more will also have
detectable iron K-shell emission lines leading to X-ray spectroscopic
redshift measurements with $\Delta z\approx0.01$ \cite{2009WP_Giacconi}.

For yet brighter clusters with $f_x=10^{-12}$~\ergscm{}, 250 objects in
total, WFXT will collect 50,000 photons per object, enough to measure
the gas temperature and metallicity profiles with the quality presently
accessible only for a handful of best-observed \emph{XMM} and
\emph{Chandra} clusters. The size and redshift reach of this sample
matches very well the \emph{REFLEX} cluster sample detected in the ROSAT
All-Sky survey. However where today we have detections with just a few
photons, we would have exceptionally detailed information for each
object from the WFXT survey.

\section{Summary}

Sensitive, wide-area X-ray surveys which would be possible with the
WFXT will detect huge samples of virialized objects spanning the mass
range from sub-groups to the most massive clusters, and extending in
redshift to beyond $z=2$. These samples will be an excellent dataset
for carrying out many traditional cosmological tests using the cluster
mass function and power spectrum. Uniquely, WFXT will be able not only
to detect clusters but also to make detailed X-ray measurements for a
large number of clusters and groups right from the survey data. Very
high quality measurements of the cluster mass function and spatial
correlation over a very wide range of masses, spatial scales, and
redshifts, will be useful for expanding the cosmological discovery
space, and in particular, in searching for departures from the
``concordant'' $\Lambda$CDM cosmological model. Finding such
departures would have far-reaching implications on our understanding
of the fundamental physics which governs the Universe

\bibliographystyle{prop}
\bibliography{wfxt}

\end{document}